\pdfoutput=1
\documentclass[aps,prd,twocolumn,superscriptaddress,preprintnumbers,floatfix,nofootinbib]{revtex4-1}

\usepackage{graphicx}
\usepackage{amsmath}
\usepackage[caption=false]{subfig}
\usepackage{siunitx}
\usepackage{placeins}
\usepackage{color}
\usepackage{standalone}
\usepackage{dcolumn}
\usepackage{tensor}
\usepackage{bm}
\usepackage{microtype}
\usepackage{etoolbox}
\usepackage{amssymb}
\usepackage{mathrsfs}
\usepackage{accents}
\usepackage[normalem]{ulem}
\usepackage[dvipsnames]{xcolor}
\usepackage[colorlinks,urlcolor=NavyBlue,citecolor=NavyBlue,linkcolor=NavyBlue,pdfusetitle]{hyperref}
\usepackage[all]{hypcap}
\usepackage[inline]{enumitem}
\usepackage[utf8]{inputenc}
\usepackage{lipsum}
\usepackage{booktabs}
\usepackage{wasysym}

\usepackage{array}

\newcommand{\ts}{\textsuperscript}

\newcommand{\beq}{\begin{equation}}
\newcommand{\eeq}{\end{equation}}

\newtoggle{commentsoff}
\togglefalse{commentsoff}

\ifdefined\nocomments
    \toggletrue{commentsoff}
\fi

\iftoggle{commentsoff}{
  \newcommand*{\mi}[1]{}
  \newcommand*{\wf}[1]{}
  \newcommand*{\comment}[1]{}
  
  \newcommand*{\todo}[1]{}
  \newcommand*{\warn}[1]{}

}{
  \newcommand*{\mi}[1]{{\color{magenta} [{\bf MAX}: #1]}}
  \newcommand*{\wf}[1]{{\color{RedOrange} [{\bf WILL}: #1]}}
  
  \newcommand*{\comment}[1]{{\color{blue} [{\bf NOTE}: #1]}}
  \newcommand*{\warn}[1]{{\color{red} [{\bf WARNING}: #1]}}
  \newcommand*{\todo}[1]{{\color{red} [{\bf TODO}: #1]}}

}

\graphicspath{{./fig/}}

\newcommand{\dcc}{LIGO-P2200028}

\begin{document}

\title{Revisiting the ringdown of GW150914}

\newcommand{\CCA}{\affiliation{Center for Computational Astrophysics, Flatiron Institute, 162 5th Ave, New York, NY 10010}}
\newcommand{\MIT}{\affiliation{LIGO Laboratory and Kavli Institute for Astrophysics and Space Research, Massachusetts Institute of Technology, Cambridge, Massachusetts 02139, USA}}
\newcommand{\STBR}{\affiliation{Department of Physics and Astronomy, Stony Brook University, Stony Brook NY 11794, USA}}
\newcommand{\CIT}{\affiliation{Department of Physics, California Institute of Technology, Pasadena, California 91125, USA}}
\newcommand{\CITLab}{\affiliation{LIGO Laboratory, California Institute of Technology, Pasadena, CA 91125, USA}}

\author{Maximiliano Isi}
\email{misi@flatironinstitute.org}
\CCA

\author{Will M. Farr}
\email{will.farr@stonybrook.edu}
\CCA
\STBR

\hypersetup{pdfauthor={Isi, Farr}}

\date{\today}

\begin{abstract}
We examine recent claims that evidence for an overtone in the ringdown of the GW150914 binary black hole merger was a result of noise anomalies. We cannot reproduce these claims, finding that our previous analysis of this event is robust to data analysis choices and consistent with the expectation that strain after the peak is well described as a superposition of quasinormal modes of the remnant black hole.
We discuss the meaning and implications of establishing that any specific ringdown mode was detected, and argue that it is misguided to expect actual LIGO-Virgo data to inform the discussion of whether or why the merger looks linear.
\end{abstract}

\maketitle

\section{Introduction}

Reference \cite{Cotesta:2022pci} revisits the analysis of the ringdown of the
black hole remnant in GW150914, claiming no evidence in favor of an overtone
mode and suggesting that claims of detection of overtone modes in that signal
\cite{Isi:2019aib} (by the present authors and collaborators) are
noise-dominated. The goal of this short paper is to (1) examine the claims in
\cite{Cotesta:2022pci} that the detection of the first overtone in the ringdown
of GW150914 was due to a noise artifact, and (2) briefly touch upon the question
of what it means to detect an individual quasinormal mode, and to what extent
such detections constrain the `linearity' of the spacetime of the remnant.

Regarding (1), we find that we cannot reproduce the key results reported
in \cite{Cotesta:2022pci}: we recover a significant non-zero overtone amplitude
stably over reasonable choices of the analysis starting time, with a decay rate
fully consistent with the expected damping of the overtone; we find that the
constraint on the deviation of the overtone frequency from the general
relativistic prediction is also stable over analysis starting time. We take this
as evidence that the results in \cite{Isi:2019aib} are robust. Regarding
(2), we argue that the question is ill-posed, with much of the focus that has
been devoted to it being misplaced. Answers to questions about the nature of the
source and whether its spacetime behaves linearly or not at a given point are
not to be found in data from gravitational wave detectors \cite{TheLIGOScientific:2014jea,TheVirgo:2014hva,KAGRA:2020tym} at the present moment.

\begin{figure*}
  \includegraphics[width=0.75\textwidth]{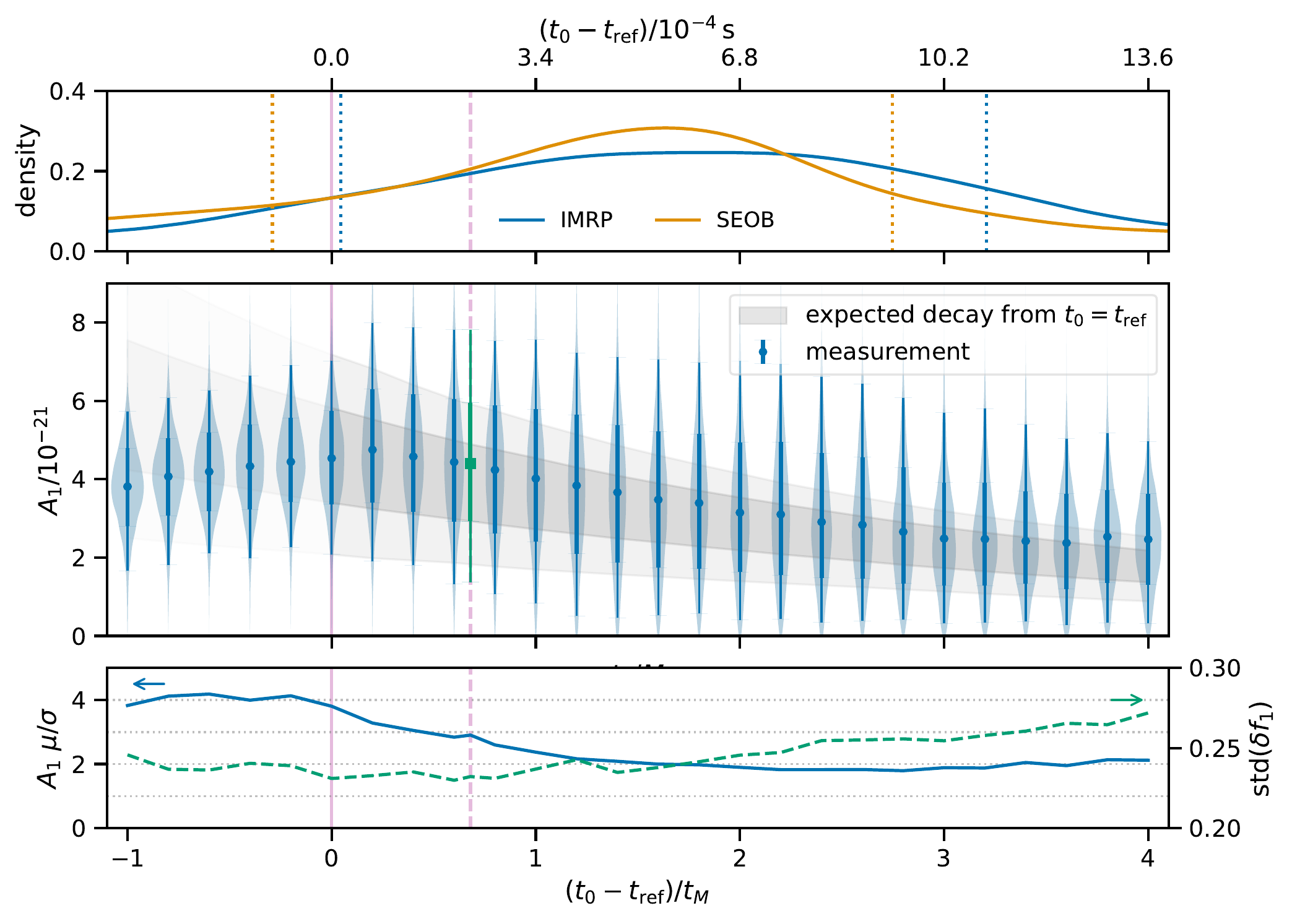}
  \caption{\label{fig:money} Analysis of the GW150914 ringdown. \emph{Top:} Posterior for the strain peak time, $\mathrm{max}(h^2_+ + h^2_\times)$, from full IMR analyses with either \textsc{IMRPhenomPv2} \cite{Husa:2015iqa,Khan:2015jqa,Schmidt:2010it} (blue) or \textsc{SEOBNRv4ROM} \cite{Taracchini:2013rva,Puerrer:2014fza,Purrer:2015tud} (orange); dotted lines enclose the 68\%-credible symmetric interval. \emph{Middle:} Posterior for the overtone amplitude $A_1$ obtained with a two-tone model ($n=0,1$) applied at different starting times $t_0$ (blue, with the preferred start time from \cite{Cotesta:2022pci} in green); the posterior is summarized by the median (circle) and 68\% (95\%) symmetric credible intervals as thick (thin) vertical lines. The dark (light) gray band shows the 68\%-credible (95\%-credible) expected decay rate starting from the measurement at $t_0=t_{\rm ref}$. \emph{Bottom:} The left axis shows the ratio of the $A_1$ posterior median to the standard deviation (blue solid curve); the right axis shows the standard deviation of the Kerr deviation in the overtone frequency ($\delta f_1$), when both $\delta f_1$ and $\delta \tau_1$ are allowed to vary within $[-0.5,0.5]$. We express time from $t_{\rm ref}$ in seconds (top axis), as well as in units of $t_M = G M /c^3$ for $M=69 M_\odot$, i.e., $t_M = 0.34\,\mathrm{ms}$ (bottom axis). As in \cite{Isi:2019aib}, the reference time corresponds to GPS time $t_{\rm ref} = 1126259462.423\, \mathrm{s}$ at the LIGO Hanford detector (vertical solid line).}
\end{figure*}

\section{GW150914 revisited}

Here we revisit the GW150914 data to examine the claims in
\cite{Cotesta:2022pci}, the core of which we understand to be: (1) the chosen
analysis start time in \cite{Isi:2019aib} was too early, (2) the detection of
the first overtone is unstable against small changes in this start time, and,
finally, (3) the overtone amplitude cannot be constrained to be nonzero except
significantly before the peak, when the model is known to be invalid.

The reference time in \cite{Isi:2019aib} (a GPS time of $t_{\rm ref} = 1126259462.423\, \mathrm{s}$ at the Hanford detector) was taken directly from the original LIGO-Virgo analysis \cite{TheLIGOScientific:2016src}, where it was claimed to be the peak time;%
\footnote{Timing posterior information for GW150914 has not been released by the
LIGO and Virgo collaborations, in spite of what is claimed in
\cite{Cotesta:2022pci}---to our knowledge, the statement in
\cite{TheLIGOScientific:2016src} was the best available at the time of writing
\cite{Isi:2019aib}. In this work, the peak time reconstructions were obtained
from our reanalysis of the full GW150914 data using \textsc{LALInference} \cite{Veitch:2014wba}.} it was chosen before carrying
out the analysis and not tuned a posteriori. The actual location of the signal
peak is uncertain because it is measured imperfectly from noisy data. It is true
that, as pointed out in \cite{Cotesta:2022pci}, the reference time in
\cite{Isi:2019aib,TheLIGOScientific:2016src} happens to fall before the median
of the reconstructed posterior; nevertheless, it lies well within the bulk of
the distribution, no more than ${\sim}1\sigma$ away from the median
(Fig.~\ref{fig:money}, top). The choice is thus adequate. 

Ringdown analyses are, for the most part, conditional on the chosen truncation
time by construction \cite{Isi:2021iql}. This is because the start time is not a
model parameter, but rather defines the data to be analyzed. The meaning and
implementation of start-time marginalization are not clear. Techniques like
those proposed in \cite{Finch:2021qph} might offer a possible avenue to allow
the start time of the ringdown analysis to vary in a controlled way.

Instead, to account for uncertainty about the arrival time of the peak of the gravitational wave strain, we can always repeat the analysis at different start times.
This was the strategy in \cite{Cotesta:2022pci}, where it was found that the posterior for the amplitude of the first overtone quickly drops to zero unless the start time is chosen to be significantly before the expected peak, where the quasinormal-mode model is known to be inapplicable.
Furthermore, \cite{Cotesta:2022pci} reported a decrease in the overtone amplitude immediately before and after the peak time used in \cite{Isi:2019aib}.
Altogether, this was interpreted as a sign that the positive identification of the overtone in \cite{Isi:2019aib} was due to a noise artifact.
However, we are unable to reproduce any of these observations.

We carry out a similar exercise as in \cite{Cotesta:2022pci} using the publicly available code in \cite{ringdown}.%
\footnote{We show results from an analysis using 0.2 s of data sampled at 2048 Hz, but our conclusions did not change when we increased the sampling rate by factors of two and four. In all cases, the identification of the data truncation---and therefore analysis start---time is done at 16 kHz, the highest available sampling rate. This is not done by the LIGO-Virgo analyses in \cite{LIGOScientific:2020tif,LIGOScientific:2021sio,LIGOScientific:2020ufj}.}
The result is summarized in Fig.~\ref{fig:money}, where the main panel shows the overtone amplitude ($A_1$) posterior obtained in a two-tone analysis (i.e., including the $n=0,1$ tones of the $\ell=m=2$ angular harmonic) starting at different times, before and after our previously chosen reference time in \cite{Isi:2019aib} ($t_0 - t_{\rm ref} = 0$ in the plot).%
\footnote{We use a strain model such that $h = F_+ h_+ + F_\times h_\times$ for detector antenna patterns $F_{+/\times}$, and polarization quadratures $h_{+} =  \frac{1}{2}(1+\cos^2\iota) \sum_{n=0,1} A_n \cos(\omega_n t + \phi_n) \exp(-t/\tau_n)$ and $h_{\times} =  \cos\iota \sum_{n=0,1} A_n \sin(\omega_n t + \phi_n) \exp(-t/\tau_n)$, where $\omega_n$ and $\tau_n$ are determined for the $n$\ts{th} overtone as a function of $M$ and $\chi$, which are themselves measured from the data together with $A_n$ and $\phi_n$. We fix the sky location to the values in \cite{Isi:2019aib}, and set the inclination to $\cos\iota=-0.99$. This corresponds to the \texttt{mchi\_aligned} model in the \textsc{ringdown} package \cite{ringdown}, modulo a factor of two in the definition of the amplitudes.}
The amplitude posterior clearly favors $A_1 >0$ for a broad range of times, without sharp fluctuations.
On the contrary, the posterior decays smoothly in full consistency with the damping expected from the first overtone; this is shown by the gray envelopes, which represent credible regions generated by drawing overtone parameters from the posterior obtained at $t_0 = t_{\rm ref}$ and projecting the decay forward in time.
This result is consistent with the true peak lying close to $t_{\rm ref}$ and the two-tone model being a good fit from that point on.

In \cite{Isi:2019aib} we reported that the $A_1$ posterior median fell $3.6\sigma$ away from zero, a measure of overtone detection significance in reference to a model with the fundamental alone, namely, $A_1 =0$ (but see \cite{Isi:2019aib,Isi:2021iql} for an account of other factors considered, including the effect of allowing for a second overtone).
We find a consistent result in our new analysis here, as shown in the bottom panel of Fig.~\ref{fig:money} ($\mu/\sigma$, blue solid curve).
For $t_0 > t_{\rm ref}$, this figure of merit decays smoothly, consistent with the evolution of the amplitude posterior.

\begin{figure}
  \includegraphics[width=\columnwidth]{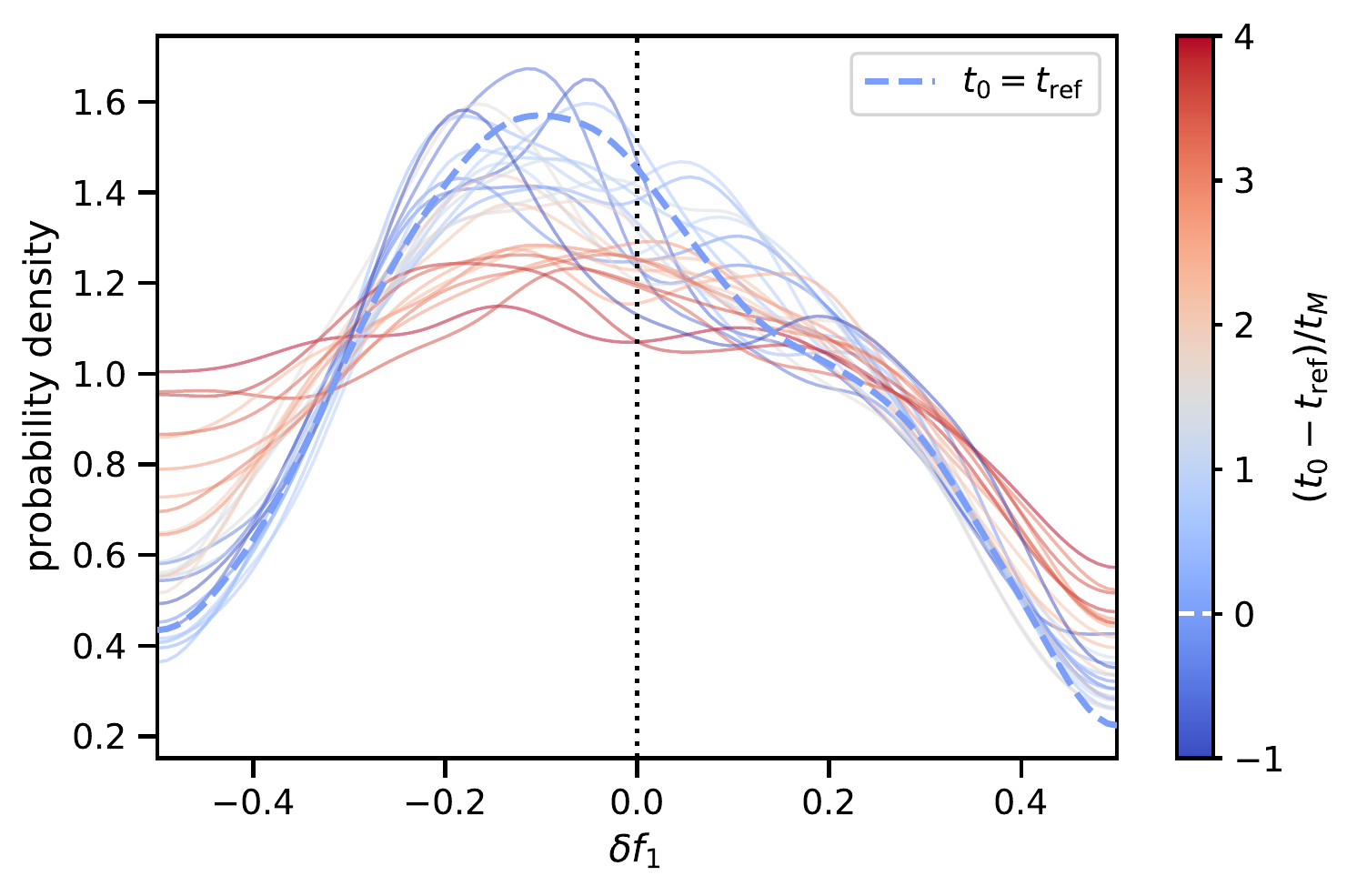}
  \caption{\label{fig:df1} Posterior for deviations from Kerr in the overtone frequency ($\delta f_1$) obtained from a two tone analysis in which both $\delta f_1$ and $\delta \tau 1$ are allowed to vary uniformly over $[-0.5, 0.5]$. The color of the trace corresponds to the analysis start time. The deviation is implemented such that $f_1 = f_1^{\rm Kerr} \exp(\delta f_1)$ \cite{Isi:2021iql}. The result for $t_0 = t_{\rm ref}$ is highlighted with a dashed trace.}
\end{figure}

\begin{figure}
  \includegraphics[width=\columnwidth]{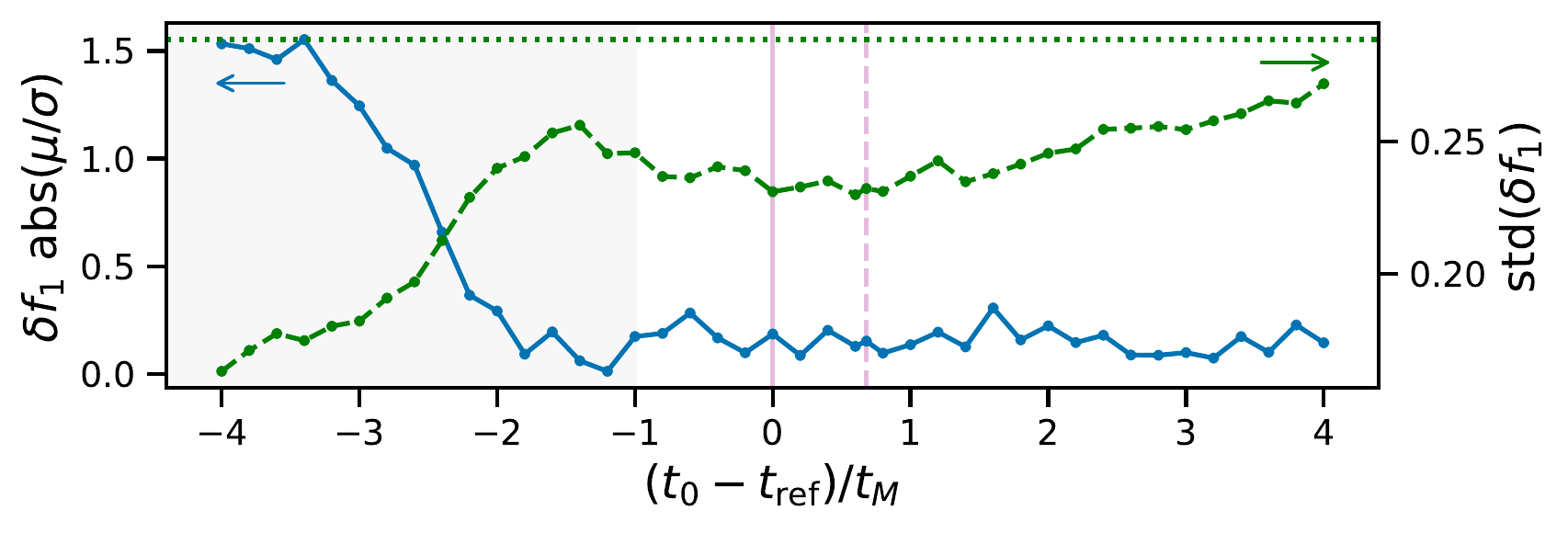}
  \caption{\label{fig:df1_std} Summary of the posteriors in Fig.~\ref{fig:df1}, plus additional analyses with start times preceding $t_{\rm ref}$ (shaded area).
  The left axis shows the absolute value of the posterior median-to-standard-deviation ratio for $\delta f_1$ (blue solid line); the right axis shows the posterior standard deviation (green dashed line), and the standard deviation of the prior (green dotted line).
  For $t_0 \lesssim t_{\rm ref} -2 t_M$, $\delta f_1$ is preferentially negative, indicating lower frequencies are favored.
  }
\end{figure}

The bottom panel of Fig.~\ref{fig:money} also shows our constraints on deviations from Kerr in the overtone frequency ($\delta f_1$) when we allow both the frequency and damping to vary around the Kerr prediction;%
\footnote{We implement the deviation such that the frequency of the overtone is $f_1 = f_1^{\rm Kerr}(M,\chi) \exp(\delta f_1)$ where $f_1^{\rm Kerr}(M,\chi)$ is the Kerr prediction for a given mass and spin, and similar for $\tau_1$. This is slightly different from the parameterization in \cite{Isi:2019aib}, where we had $f_1 = f_1^{\rm Kerr}(M,\chi) \left(1+\delta f_1\right)$; we discuss this in \cite{Isi:2021iql}.}
the constraint is summarized by the $\delta f_1$ posterior standard deviation ($\mathrm{std}(\delta f_1)$, green dashed curve).
As we expect, the frequency is best constrained for start times around $t_{\rm ref}$.
For later start times, the posterior support for $A_1=0$ increases and the frequency is more poorly constrained; this is clear also from the full $\delta f_1$ posteriors in Fig.~\ref{fig:df1}.
On the other hand, including data significantly before $t_{\rm ref}$ can cause the model to prefer deviations away from the Kerr spectrum, indicating that Kerr quasinormal modes are not a good model for the data at that point---the transition between the two regimes occurs sharply around $t_0 \approx -1.5 t_M$ (Fig.~\ref{fig:df1_std}).

\begin{figure}
  \includegraphics[width=\columnwidth]{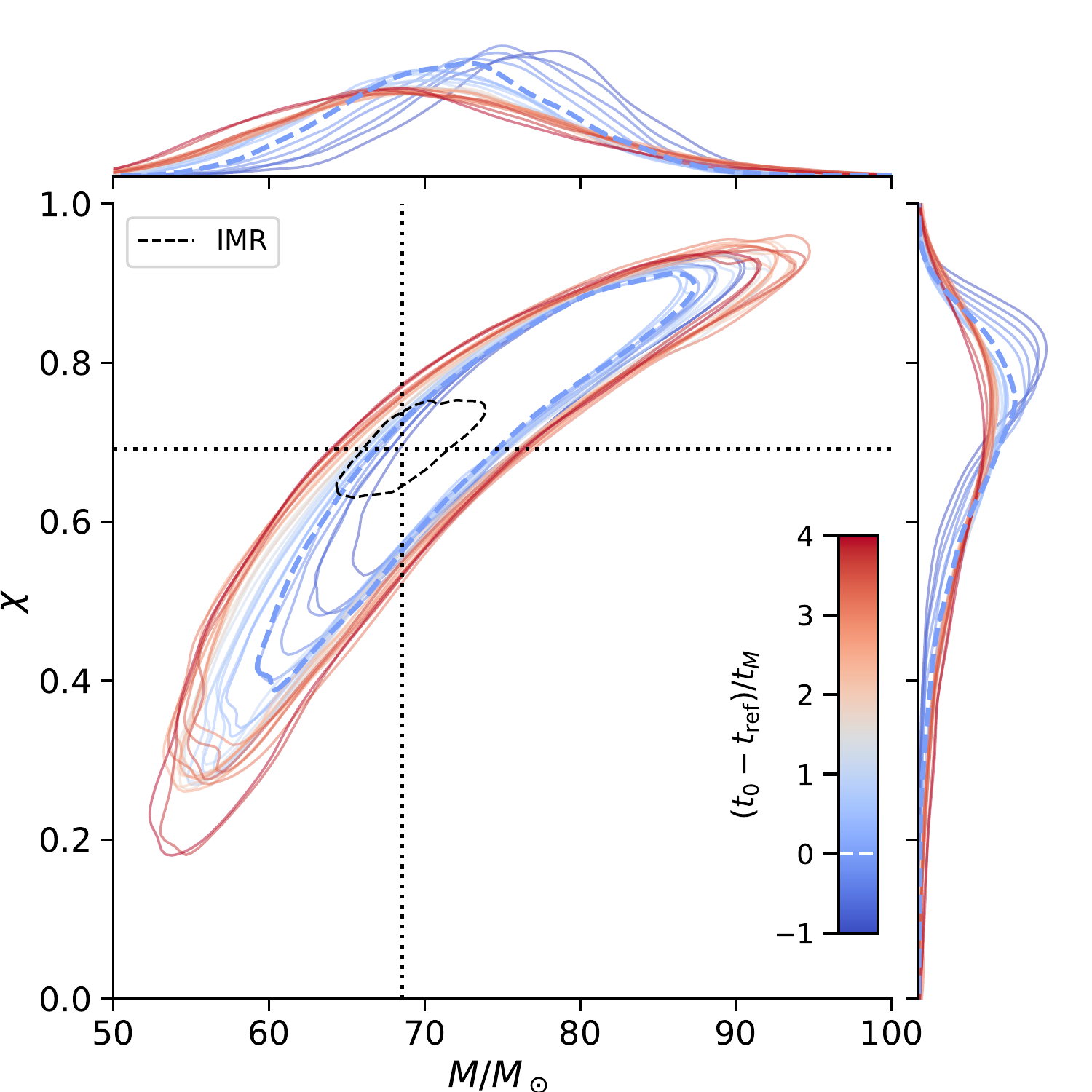}
  \caption{\label{fig:mchi_n1} Posterior for the remnant mass ($M$ in solar masses $M_\odot$) and dimensionless spin magnitude ($\chi$) obtained with a two-tone analysis ($n=0,1$) beginning at different start times (color). Contours enclose 90\% of the probability mass. The result for $t_0 = t_{\rm ref}$ is highlighted with a dashed blue trace. The result from the IMR analysis is also shown for comparison (dashed black), and was obtained from the LIGO-Virgo samples \cite{GWOSC:GWTC1,LIGOScientific:2019lzm} as detailed in \cite{Isi:2019aib}. }
\end{figure}

\begin{figure}
  \includegraphics[width=\columnwidth]{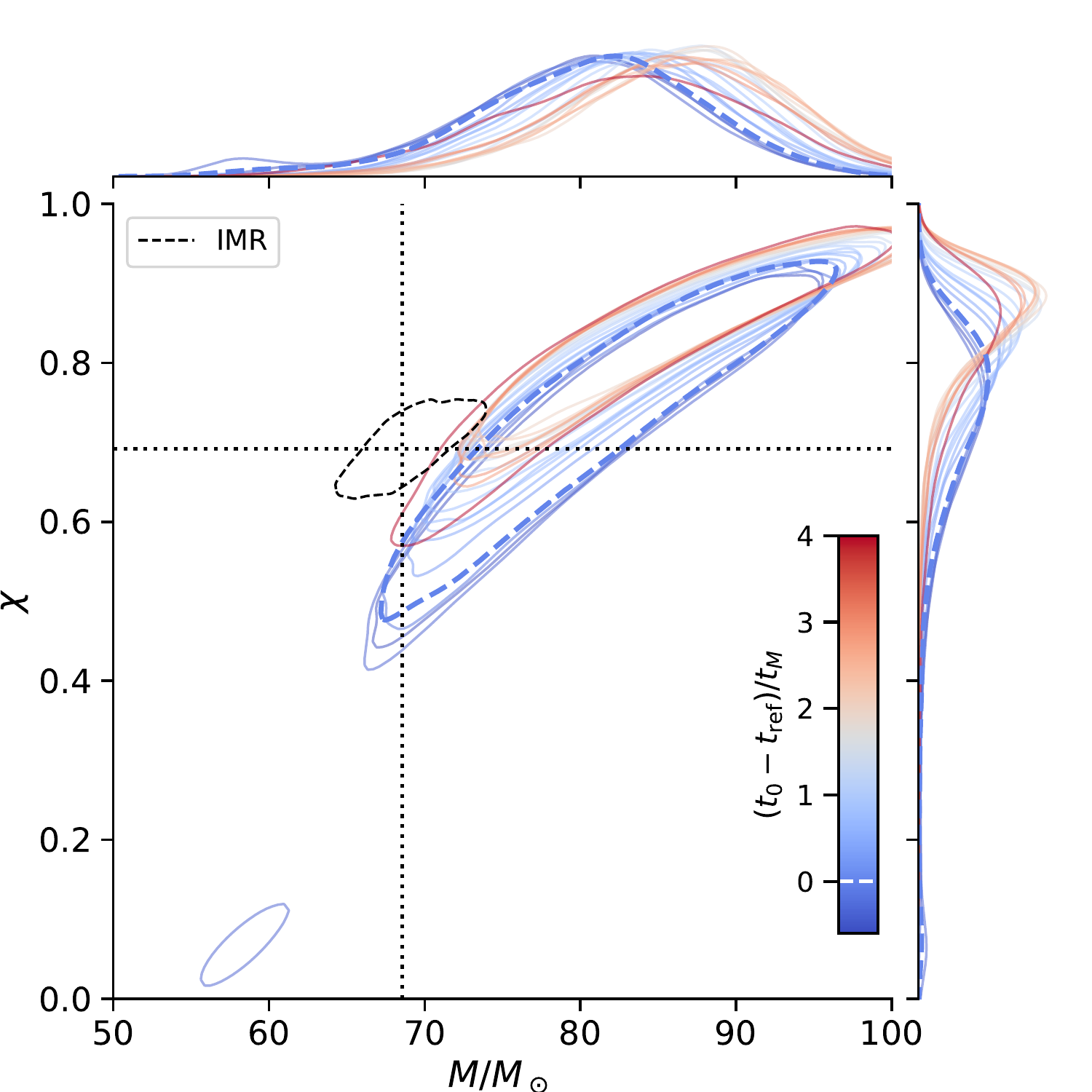}
  \caption{\label{fig:mchi_n0} Same as in Fig.~\ref{fig:mchi_n0} but for a model including just the fundamental mode ($n=0$). This model is a poor description of the data at these early times.}
\end{figure}

A similar conclusion can be drawn from the posterior for the remnant mass ($M$) and dimensionless spin magnitude ($\chi$) obtained assuming a Kerr spectrum in the analyses of Fig.~\ref{fig:money}.
As shown in Fig.~\ref{fig:mchi_n1}, the remnant posterior stabilizes around $t_0 \approx t_{\rm ref}$, at which point it begins to show better consistency with the expectation from the full inspiral-merger-ringdown analysis (IMR, black dashed).
This is not true for a model including the fundamental mode alone (Fig.~\ref{fig:mchi_n0}), which fails to agree with the IMR analysis at these early times.

Besides looking at GW150914 itself, we also inject a number of simulated signals in real data around the event (Fig.~\ref{fig:inj}).
We find results consistent with those above: the overtone amplitude posterior varies as might be expected from regular noise fluctuations, and nothing indicates that nonstationarities in the data are derailing the analysis.
This experiment is meant to account for non-Gaussianities in actual detector noise; injections in Gaussian noise were already presented in \cite{Isi:2021iql}.

Our results here and in \cite{Isi:2019aib}, like in \emph{any} analysis, are predicated on a number of choices about data conditioning and noise power-spectral-density (equivalently, autocovariance-function) estimation.
For different reasonable choices for these settings, we obtain slightly different results, but this variation is within expectation and do not alter our conclusions.

\section{The meaning and implications of detecting a mode}

The problem of establishing whether a specific ringdown mode has been detected is made difficult by the fact that the meaning of ``detection'' is not itself clear in this context.
In the regime where we can inarguably expect a superposition of quasinormal modes to be a good description of the data, one can reasonably phrase this question as a choice between $A_1 = 0$ and $A_1 > 0$ within a model including both $n=0,1$---that is, presuming no other quasinormal modes could be expected to be relevant.
Already in that simplified scenario the question is technically ill-posed: under that model, the probability that $A_1 = 0$ exactly is effectively zero; this problem appears in other contexts, e.g., the determination of whether a binary black hole system is precessing or whether a given angular harmonic has been detected.%
\footnote{Lack of precession requires the individual black-hole spins to be \emph{exactly} perpendicular to the orbital plane, which is practically impossible to achieve physically.
Higher-than-quadrupole multipoles of the radiation are technically present in all realistic signals, even if their amplitude is minimal; furthermore, models that simply ignore them while keeping the quadrupole as in general relativity are unphysical. See, e.g., \cite{GW190412,LIGOScientific:2020ufj} for relevant discussions.}
The problem is made even murkier if, like in \cite{Cotesta:2022pci}, one allows for the possibility that data are not a superposition of damped sinusoids and neither of the $n=0$ nor the $n=0,1$ models are valid.
In that case, there is no reason to expect a preference for either model to be informative (whether quantified via the amplitude posterior or a Bayes factor).

Bayes factors are an especially poor tool for this purpose; the drawbacks of Bayes factors in the context of this problem were already discussed in \cite{Isi:2021iql}, so we do not repeat that here.
Nevertheless, Fig.~\ref{fig:bf} shows the Bayes factors corresponding to the amplitude posteriors in Fig.~\ref{fig:money}; the Bayes factors favor the presence of the overtone even for $t_0 > t_{\rm ref}$, unlike in \cite{Cotesta:2022pci}.
Of course, as remarked in \cite{Isi:2021iql}, the overall vertical location of the Bayes factor curve can be shifted arbitrarily by modifying the maximum allowed value of $A_1$ in the prior (in our case $A_1 = 10^{-20}$).
Besdies prior differences, comparing Bayes factors from different analyses can also be inappropriate when the models used are not identical---for that reason comparisons to Bayes factors in \cite{LIGOScientific:2020tif} are not necessarily interesting or valid.

\begin{figure}
  \includegraphics[width=\columnwidth]{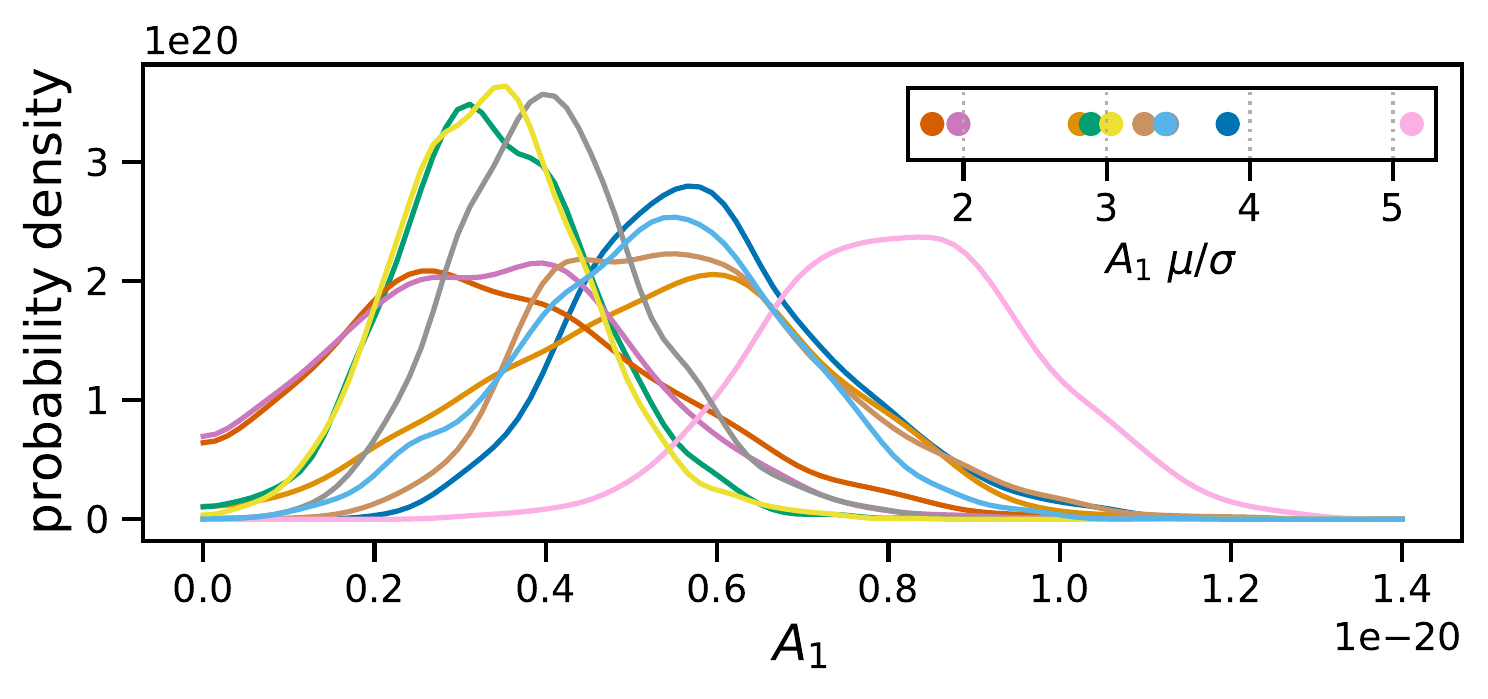}
  \caption{\label{fig:inj} Posterior for the overtone amplitude $A_1$ obtained for injections of the GW150914-like numerical simulation \texttt{SXS:BBH:0305} \cite{SXSwf,Boyle:2019kee,Schmidt:2017btt} into actual LIGO noise around GW150914. We set the total redshifted mass to be $M_{\rm tot}=72\, M_\odot$, with a luminosity distance of 450 Mpc. The inset shows the value of the median-to-standard-deviation ratio. The variation in the amplitude posterior agrees with what we expect from Gaussian noise.} 
\end{figure}

\begin{figure}
  \includegraphics[width=\columnwidth]{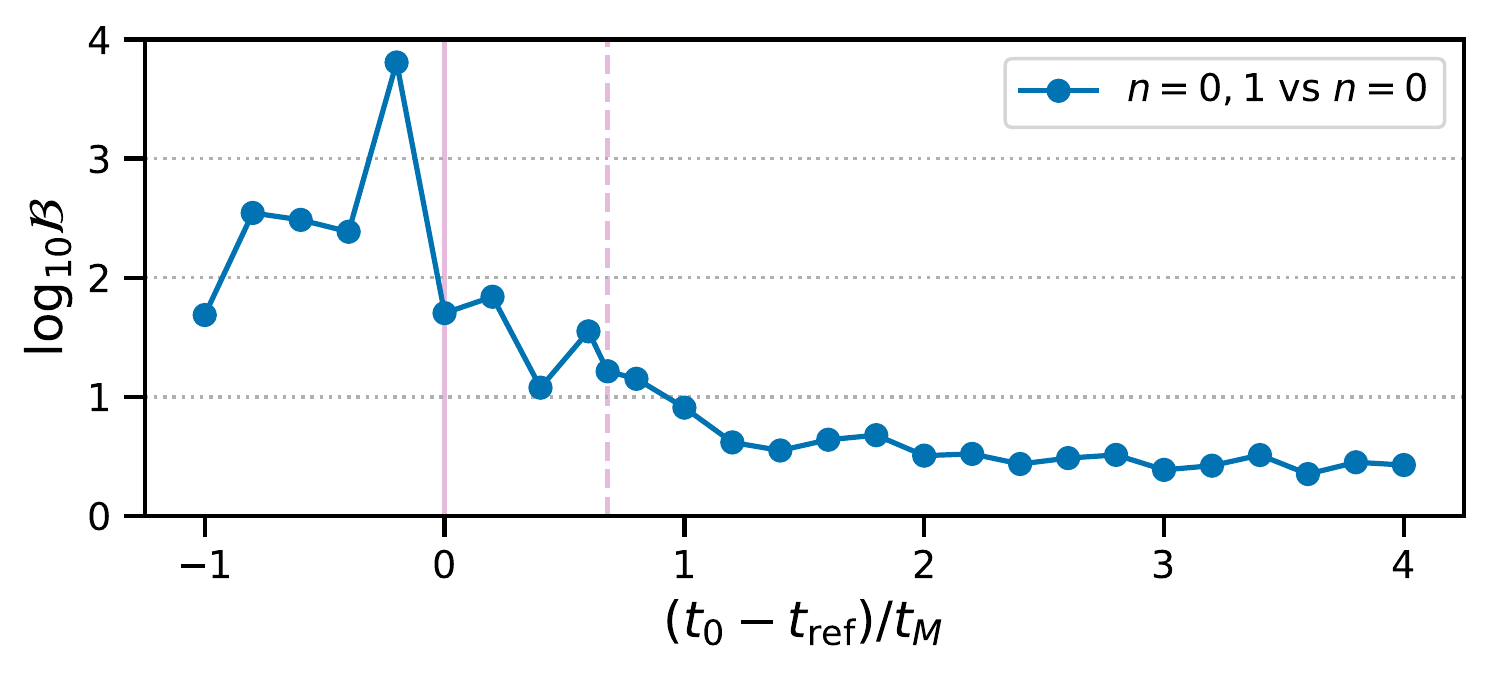}
  \caption{\label{fig:bf} Base-ten logarithm of Bayes factors comparing a two-tone model ($n=0,1$) to a one-tone model ($n=0$) for runs presented in Fig.~\ref{fig:money}. The values are computed from the amplitude posterior via the Savage-Dickey density ratio, for a prior uniform in $0\leq A_1 \leq 10^{-20}$.} 
\end{figure}

In any case, the question of whether a mode was detected is inherently
``fuzzy.'' To quote \cite{gelman2006difference}: the difference between
statistically significant and not statistically significant is not, itself,
statistically significant.  This can be seen most clearly in Figure
\ref{fig:inj}, where a signal of constant amplitude injected into noise with
(presumably) identical statistical properties generates, on average, a ${\sim} 3.5
\sigma$ overtone ``detection;'' but with fluctuations $\sim \pm 1 \sigma$, as
expected from the intrinsic variation of the particular noise realization from
moment to moment.  Thus the $3.6\sigma$ ``detection'' in \cite{Isi:2019aib}
could just as easily have been a marginal $2\sigma$ ``suggestion'' or a
$5\sigma$ ``confident detection'' of the overtone's presence.  
Compounded with the above, this makes the question of ``how significant is the
detection of an overtone'' not particularly interesting.

In the end, much of the interest in whether the overtone has positively been detected in GW150914 seems to be fueled by a wish to empirically test the results in \cite{Giesler:2019uxc} that the gravitational waves from a binary black hole merger can be described as a superposition of quasinormal modes of the remnant black hole from the peak of the strain onward; however, this urge is misguided.
As made clear in \cite{Isi:2019aib}, ringdown analyses starting at the peak \emph{presume} the results of \cite{Giesler:2019uxc} to hold, and are not designed to corroborate them (at least not directly).

As we have argued elsewhere, all that is needed before confronting the data in an analysis like \cite{Isi:2019aib} is the \emph{empirical} confirmation from numerical relativity studies that, starting from the peak of the strain, the gravitational-wave emission from a binary black hole should be well described by a superposition of quasinormal modes \emph{with frequencies and damping rates of the remnant black hole} (and not others).
This is indeed the conclusion of \cite{Giesler:2019uxc} and other studies (particularly Fig.~7 in \cite{Giesler:2019uxc}).
If, after analyzing real data, the quasinormal modes are not found to be consistent with a final Kerr remnant, then the observation does not agree with what we expect from the merger of two Kerr black holes in general relativity.\footnote{Or, like in any other analysis, there is a failure of at least one of the many underlying assumptions---including that the noise is stationary, that its spectrum has been correctly estimated, and that the analysis start time has been properly chosen. See \cite{LIGOScientific:2019fpa,LIGOScientific:2020tif} for relevant discussions of the role of systematics in tests of general relativity with gravitational waves.}

The ability to carry out such a cross-check is precisely the reason we are interested in identifying multiple quasinormal modes in the first place.
To the extent that we can carry out this Kerr-consistency test, the mode was detected.
Concretely, as we argued in \cite{Isi:2019aib}, the $\delta f_1$ (or $\delta\tau_1$) posterior can only be informative insofar as the data support a nonzero overtone amplitude.
If that is the case, we can say the overtone was ``detected'' and check for one of two scenarios: (1) the deviation posterior is consistent with zero and thus also with Kerr black holes in general relativity, or (2) the deviation posterior disfavors zero, indicating that one or more of the assumptions in the model is broken (including, potentially, the assumption that general relativity is correct).

\section{Conclusion}

We have revisited the ringdown of GW150914 finding our previous claims in
\cite{Isi:2019aib} to be robust against choices in analysis start time, within
the expected location of the signal peak from inspiral-merger-ringdown analyses.
We cannot reproduce the claims to the contrary presented in
\cite{Cotesta:2022pci}. The robustness of our observations against choices of
sampling rate and power-spectrum estimation methods gives us additional
confidence in our results. We make code to reproduce results in this paper
available in \cite{release}.

\begin{acknowledgments}
The Flatiron Institute is a division of the Simons Foundation, supported through the generosity of Marilyn and Jim Simons.
This material is based upon work supported by NSF's LIGO Laboratory which is a major facility fully funded by the National Science Foundation.
This research has made use of data or software obtained from the Gravitational Wave Open Science Center (gw-openscience.org), a service of LIGO Laboratory, the LIGO Scientific Collaboration, the Virgo Collaboration, and KAGRA. LIGO Laboratory and Advanced LIGO are funded by the United States National Science Foundation (NSF) as well as the Science and Technology Facilities Council (STFC) of the United Kingdom, the Max-Planck-Society (MPS), and the State of Niedersachsen/Germany for support of the construction of Advanced LIGO and construction and operation of the GEO600 detector. Additional support for Advanced LIGO was provided by the Australian Research Council. Virgo is funded, through the European Gravitational Observatory (EGO), by the French Centre National de Recherche Scientifique (CNRS), the Italian Istituto Nazionale di Fisica Nucleare (INFN) and the Dutch Nikhef, with contributions by institutions from Belgium, Germany, Greece, Hungary, Ireland, Japan, Monaco, Poland, Portugal, Spain. The construction and operation of KAGRA are funded by Ministry of Education, Culture, Sports, Science and Technology (MEXT), and Japan Society for the Promotion of Science (JSPS), National Research Foundation (NRF) and Ministry of Science and ICT (MSIT) in Korea, Academia Sinica (AS) and the Ministry of Science and Technology (MoST) in Taiwan.
This paper carries LIGO document number \dcc{}.
\end{acknowledgments}

\bibliography{refs}

\end{document}